\documentclass[12pt]{article}
\usepackage{amsfonts}
\usepackage{hyperref}
\usepackage{amsmath}
\usepackage{epsfig}

\newcommand{\eq}{\begin{equation}}
\newcommand{\eqx}{\end{equation}}
\newcommand{\eqs}{\begin{equation*}}
\newcommand{\eqsx}{\end{equation*}}
\newcommand{\eqn}{\begin{eqnarray}}
\newcommand{\eqnx}{\end{eqnarray}}
\newcommand{\alg}{\begin{align}}
\newcommand{\algx}{\end{align}}
\newcommand{\ttau}{\tilde{\tau}}     
\def\d{\partial}

\newcommand{\f}[2]{\frac{#1}{#2}}

\newcommand{\nn}{{\cal N}}
\newcommand{\zt}{\tilde{z}}

\newcommand{\sg}{\sigma}

\newcommand{\cor}[1]{\left\langle{#1}\right\rangle}
\newcommand{\rsq}{{\mathfrak R^2}}
\newcommand{\eps}{\varepsilon}

\begin{document}
\title{Hydrodynamic Flow of the Quark-Gluon Plasma and Gauge/Gravity 
Correspondence\thanks{$48^{th}$ Cracow School of Theoretical Physics: Aspects of 
Duality, 
 June 13-22, 2008     Zakopane, Poland.}} %

\author{Michal P. Heller$^{a}$, Romuald A. Janik$^{a}$ and R. 
Peschanski$^b$\thanks{
{\tt heller@th.if.uj.edu.pl},~{\tt ufrjanik@th.if.uj.edu.pl}, {\tt 
robi.peschanski@cea.fr}}
 \vspace{.1cm}\\
$^a$ \small M.Smoluchowski Institute of Physics,\\ 
\small Jagellonian University,
\small Reymonta 4, 30-059 Krakow, Poland. \\
$^b$ \small Institut de Physique Th{\'e}orique \\
\small URA 2306, Unit\'{e} de Recherche associ{\'e}e au CNRS \\
\small CEA-Saclay, F-91191 Gif/Yvette Cedex, France.}
\date{}
\maketitle
  
\begin{abstract}
The contribution presents a summary of the Gauge/Gravity approach to the study 
of hydrodynamic flow of the quark-gluon plasma formed in heavy-ion collisions. 
Considering the ideal case of a supersymmetric Yang-Mills theory for which the 
AdS/CFT correspondence gives a precise form of the Gauge/Gravity duality, the 
properties of the strongly coupled expanding plasma are put in one-to-one 
correspondence with the metric of a 5-dimensional black hole moving away in the 
5th dimension and its
deformations consistent with the relevant Einstein equations. Several recently 
studied aspects of this framework are recalled and put in perspective.  This 
paper is a written version of the four lectures given by the authors on 
that subject.
\end{abstract}
\vfill
\pagebreak

\section{Hydrodynamics are relevant for heavy-ion collisions}

One of the most striking lessons one may draw \cite {review,hydro} from  
experiments on  
heavy-ion collisions at high energy ($e.g.$ at the RHIC accelerator, Brookhaven) 
is 
that fluid hydrodynamics seems to be relevant for understanding the dynamics of 
the reaction.  Indeed, the elliptic flow 
\cite{JY} describing the anisotropy of the low-p$_T$ particles produced in  
a collision at non zero impact parameter implies the existence of a collective 
flow of the particles following a hydrodynamical  pressure gradient due to the 
initial eccentricity in the collision. Moreover most hydrodynamical 
simulations which are successful to describe this elliptic flow are consistent 
with an almost ``perfect fluid'' behaviour, $i.e.$ a small ``viscosity over 
entropy'' ratio $\eta/s$ (see, for instance, the reviews \cite{hydro}).

\begin{figure}[ht]
\begin{center}
\includegraphics[width=30pc]{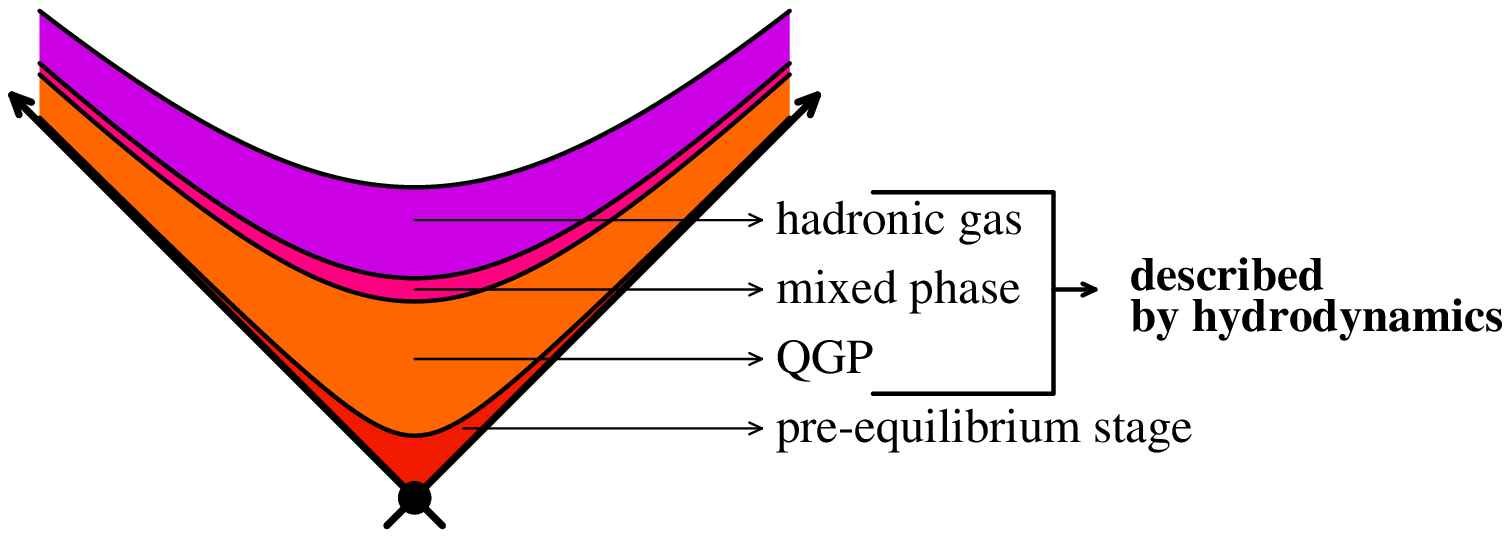}
\end{center}
\caption{\small{\it Description of QGP formation in heavy ion collisions}. The 
kinematic landscape is defined by ${\tau = \sqrt{x_0^2-x_1^2}\ ;\ {\eta=\f 12 
\log \f 
{x_0+x_1}{x_0-x_1}}\ ;\ {x_T\!=\!\{x_2,x_3\}}}\ ,$
where the coordinates along the light-cone are $x_0 \pm x_1,$ the transverse 
ones 
are $\{x_2,x_3\}$ and $\tau$ is the proper time, $\eta$ the ``space-time 
rapidity''.}
\label{1}
\end{figure}

The validity of a hydrodynamical description assuming a 
quasi-perfect fluid behaviour has been nicely anticipated  in 
Ref.\cite{Bjorken}. The so-called {\it Bjorken flow} is based on the hypothesis 
of an intermediate 
stage of the reaction process, namely a boost-invariant\footnote{The 
introduction of hydrodynamics in the 
description of  high-energy 
hadronic collisions has been proposed by Landau \cite{landau}, assuming  ``full 
stopping'' initial conditions which result in a non boost-invariant solution or 
$Landau\ flow$ (see \cite{us} for a unified description of Bjorken and Landau 
flows). We will comment later on the relevance of the Landau flow for AdS/CFT.}
 quark-gluon plasma phase as a  relativistic expanding fluid. It is  formed  
after a (quite rapid) thermalization period and finally decays into hadrons, see 
Fig.~\ref{1}. The boost-invariance can be justified  in the central region 
of the collision since the observed distribution of particles is flat, in 
agreement with the prediction of hydrodynamical boost-invariance, where 
(space-time) fluid 
and (energy-momentum) particle rapidities are proved to be equal \cite{Bjorken}, 
see section 3.

The Bjorken flow was instrumental for deriving many qualitative and even  
quantitative 
features of   the quark-gluon plasma formation in heavy-ion 
reactions. However, as inherent to the hydrodynamic approach, it says only 
little on the relation with the microscopic gauge field theory, $i.e.$ Quantum 
Chromodynamics (QCD). Some 
important questions  remain unsolved, such as  the reason why the fluid 
behaves like a perfect fluid, what is the small amount of viscosity it may 
require,  why and how fast thermalization proceeds, etc... The problem is made 
even 
more difficult by the  strong coupling regime of QCD which is very probably 
required, since a perturbative description leads in general to a high  $\eta/s$. 
Indeed,  the  mean free path induced by the gauge theory should be small (hence 
the coupling strong) in order to damp  the near-by force  transversal to the 
flow, measuring the shear viscosity. 

It is thus interesting to use our modern (but still largely in progress) 
knowledge
of non perturbative methods in quantum field theory to fill the gap between the 
macroscopic and microscopic descriptions of the quark-gluon plasma produced in 
heavy-ion collisions. Lattice gauge theory methods are very useful to analyze 
the static properties of the quark-gluon plasma, but there are still powerless 
to describe the plasma in collision. Hence we are led to rely upon the new 
tools offered by the Gauge/Gravity correspondence and in particular the one 
which is the most studied and well-known namely the AdS/CFT duality 
\cite{adscft} between the 
$\nn=4$ supersymmetric Yang-Mills theory and the type IIB superstring in the 
large $N_c$ approximation. The features of the gauge theory on the (physical) 
Minkowski space in $3+1$ dimensions at strong coupling are in one-to-one 
relation with  corresponding ones in the bulk of the target space of the 10-d 
string and in particular in the 5-dimensional metric of the AdS space, the boundary of which 
can be identified with the 4-dimensional Minkowski space.

One should be aware when using the AdS/CFT tools that there does not yet exist a 
gravity dual construction 
for QCD. However, the nice feature of the quark-gluon plasma problems is that it 
is 
a deconfined phase of QCD, characterized by collective degrees of freedom  and 
thus one may expect to get useful information from  AdS/CFT duality. This has 
been already proved when describing static geometries by an 
evaluation of  $\eta/s$  \cite{son}.  The subject of the present lectures is the 
investigation of the 
Gauge/Gravity correspondence, in particular the AdS/CFT duality, in a dynamical 
setting corresponding to a 
collision. 

\section{{\it Relativistic Hydrodynamics} and  {\it Bjorken Flow}}
\label{sechydro}

On theoretical grounds, there are quite appealing features for
applying hydrodynamic concepts to high-energy heavy-ion reactions. Such
concepts have been already introduced some time ago \cite{landau,Bjorken} and 
find a plausible 
realization
nowadays. The fact that a rather dense interacting medium is created in the 
first stage of
the collision allows one to admit that the individual partonic or
hadronic degrees of freedom are not relevant during the early evolution
of the medium and justifies its treatment as a fluid. For the same reason local 
equilibrium is a plausible assumption. Moreover, the high
quantum occupation numbers allow one to use a classical picture and to
assume that the ``pieces of fluid'' may follow quasi-classical
trajectories in space-time, expressed as an in-out cascade
\cite{gl} with straight-line trajectories starting at the origin (see Fig. 2), with
\eq
y=\eta \label{bgl}
\eqx
where
\eq
y=\frac12 \log\left(\frac{E+p}{E-p}\right) \;;\;\;
\eta=\frac12 \log\left(\frac{x_0+x_1}{x_0-x_1}\right) 
\eqx
are respectively the rapidity and ``space-time rapidity'' of the piece of the 
fluid\footnote{We keep the conventional notation $\eta$, not to be confused with 
 viscosity. The difference is clear enough to avoid mistakes.}.
\begin{center}
\begin{figure}[ht]
$\ \ \ \ \ \ \ \ \ \ \ \ \ \ \ \ \ \ \ \ \ 
$\includegraphics[width=8cm]{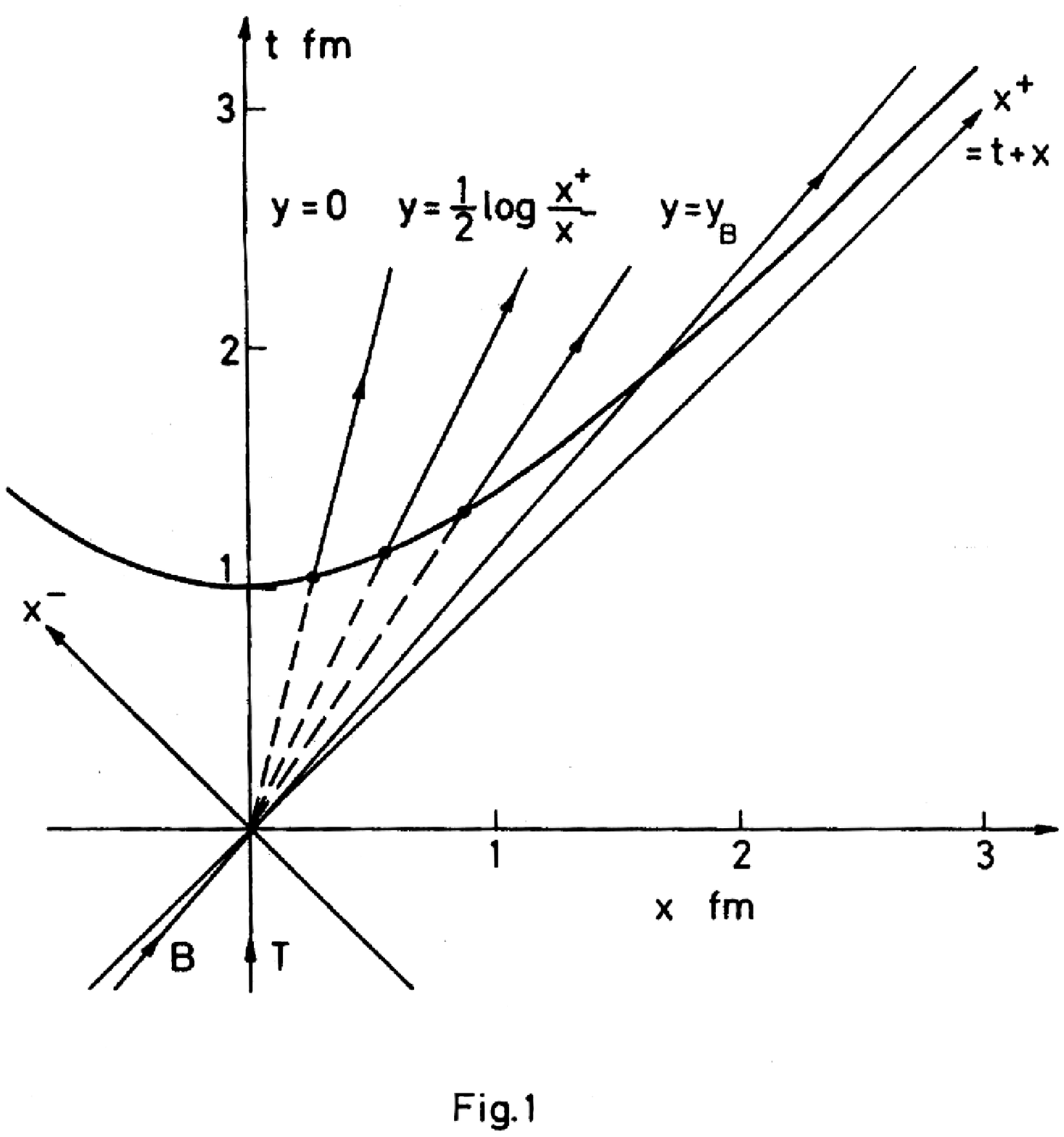}
\label{2}
\caption{{\it In-Out cascade.} The ``piece of fluid'' with space-time rapidity 
$\eta$ gives rise to hadrons at rapidity $y\equiv\eta,$ after crossing the 
``freeze-out'' hyperbola at fixed proper-time $\tau.$}\end{figure} 
\end{center}     
Note that (\ref{bgl}) can be rewritten in the form
\eq
2y=\log u^+-\log u^-= \log x^+- \log x^- \label{uz}
\eqx where
$u^\pm=e^{\pm y}$ are the light-cone components of the fluid
(four-)velocity and $x^\pm=x_0\pm x_1$ are the light-cone kinematical     
variables.      
     
Taking (\ref{bgl}) as the starting point and using the perfect fluid     
hydrodynamics, Bjorken developped in his seminal paper \cite{Bjorken} 
a suggestive (and
very useful in many applications) physical picture of the 
central     
rapidity region of highly relativistic collisions of heavy ions. In this picture 
the 
condition (\ref{bgl}) leads to a     boost-invariant geometry of the expanding 
fluid 
and thus to the central
plateau in the distribution of particles.     
     
Let us introduce the relativistic hydrodynamic equations in light-cone 
variables. We consider the ``perfect fluid'' approximation for which the 
energy-momentum tensor is
\eq    
 T^{\mu\nu}= (\epsilon+p)u^{\mu}u^{\nu} - p \eta^{\mu\nu}  \label{T}
\eqx     
where $\epsilon$ is the energy density, $p$ is the pressure and      
$u^{\mu}$ is the 4-velocity. We assume that the energy density and pressure
 are related      
by the equation of state:     
\eq     
\epsilon = gp 
\label{state}     
\eqx     
 where $1/\sqrt{g}$ is the sound velocity in the liquid. For the ``conformal 
case'' $ T^{\mu\mu}=0$ and thus $g\equiv 3.$  
    
Using     
\eq     
u^\pm\equiv u^0\pm u^1=e^{\pm y}     
\eqx
and introducing     
\eq     
x^\pm=  x^0\pm x^1 =\tau e^{\pm \eta}\; \rightarrow \;
(\frac {\d}{\d x^0}\pm \f {\d}{\d x^1})={\scriptstyle\ \frac12}  \f {\d}{\d 
x^\pm}   
\equiv{ \scriptstyle\ \frac12} \d_\pm     
\eqx     
where $\tau=\sqrt{x^+x^-}$ is the proper time and $\eta$ is the spatial     
rapidity of the fluid,      
the hydrodynamic equations      
     \eq     
\d_\mu T^{\mu\nu} =0     
\eqx     
take the form     
\eq     
\d_\pm T^{01}+\frac12\d_+(T^{11}\pm T^{00})     
-\frac12 \d_-(T^{11}\mp T^{00})=0\ .     
\eqx     
Using now (\ref{T}) and the equation of state (\ref{state})     
 we deduce from this     
\eqn     
g\d_+[\log p]&=& -\frac{ (1\!+\!g)    ^2}2\d_+y-\frac{g^2\!-\!1   
}2e^{-2y}\d_-y \nonumber\\     
g\d_-[\log p]&=&\frac{ (1\!+\!g)    ^2}2\d_-y +\frac{g^2\!-\!1   }2e^{2y}\d_+y \ 
. 
\label{ef}     
\eqnx     
These are two equations for two unknowns which describe     
 the state of the liquid: the pressure $p$ and the     
rapidity $y$. They should be expressed in terms of the     
positions $x^+,x^-$ in the liquid. Other thermodynamic quantities can be     
obtained from the equation of state (\ref{state}) and the standard     
thermodynamical identities:     
\eq      
p+\epsilon = Ts\;;\;\; d\epsilon = T ds \label{therm}     
\eqx     
where we have assumed for simplicity vanishing chemical potential.     
     
The result is      
\eq     
\epsilon =gp=\epsilon_0 T^{g+1}\;;\;\;s=s_0T^g\;\rightarrow \; s\sim     
\epsilon^{g/(g+1)} . \label{esp}     
\eqx 
     
The simplest possibility to describe the expansion of the fluid     
was suggested by Bjorken \cite{Bjorken} who proposed to use the Ansatz 
(\ref{bgl}) 
in the hydrodynamical context.
Introducing (\ref{bgl}) into (\ref{ef}) we obtain      
\eq     
g\d_+[\log p]=-\frac{1+g}{2x^+}\;;\;\;     
g\d_-[\log p]=-\frac{g+1}{2x^-}     
\eqx     
from which we deduce      
\eq     
p=\epsilon\ g^{-1}= p_0\ (x^+x^-)^{-(g+1)/2g} = p_0\ \tau^{-(g+1)/g}\ ,     
\eqx     
where $p_0$ is a constant, and thus specifically
    
\eq     
p=\epsilon/3= p_0\ (x^+x^-)^{-2/3} = p_0\ \tau^{-4/3} \propto   T^4    
\label{g3}\eqx     
for the conformal case.

Thus the system is     
boost-invariant: the pressure     
does not depend neither on $\eta$ nor on $y$. So are $\epsilon$, $s$ and $T$,    
 given by (\ref{esp}). 
 It is interesting to note that the Landau flow  corresponds asymptotically only 
to a logarithmic correction of relation \eqref{bgl}, namely
     
\eq
u^{\pm}\sim x^{\pm}\sqrt{\log x^{\pm}} \ ,\label{l}
\eqx
 which gives finally rise (as already noticed in \cite{landau}, and for instance 
 recently discussed in \cite{us1}) to a gaussian shape in the $y$ distribution 
of the entropy, revealing a non boost-invariant picture, at least at some 
distance from central rapidity.
    
\section{Interest of AdS/QCD duality}

 In the previous sections  we mentionned the ubiquity of hydrodynamic methods in 
the description of QGP produced at RHIC. Yet, despite their success in 
describing data, we have to keep in mind that they are used as a 
phenomenological model without a real derivation from gauge theory. This is 
quite understandable since almost perfect fluid hydrodynamics is intrinsically a 
strong coupling phenomenon - for which one lacks a purely gauge theoretical 
method\footnote{Lattice QCD methods do not work well here as this would require 
analytical continuation to Minkowski signature which is nontrivial in this 
context}. 

On the other hand there exists a wide class of gauge theories, which can be 
studied analytically at strong coupling. These are superconformal field theories 
with gravity duals. String theory methods (namely AdS/CFT correspondence) maps 
gauge theory dynamics (CFT) at strong coupling and large number of colors into 
solving Einstein equations in asymptotically anti-de Sitter space (AdS). The 
theories with gravity duals can differ substantially from real world QCD at zero 
temperature. The best known example of such theory - $\mathcal{N} = 4$ super 
Yang-Mills (SYM) - is a superconformal field theory with matter in the adjoint 
representation of the gauge group SU(N$_{c}$). Because of the conformal symmetry 
at the quantum level this theory does not exhibit confinement. On the other hand 
differences between $\mathcal{N} = 4$ SYM and QCD are less significant above 
QCD's critical temperature, when quarks and gluons are in the deconfined phase. 
Moreover it was observed on the lattice that QCD exhibits a quasi-conformal 
window in the certain range of temperatures, where the equation of state is 
well-approximated by $\epsilon = 3 p$. The above observations together with 
experimental results suggesting that quark-gluon plasma is a strongly coupled 
medium is an incentive to use the AdS/CFT correspondence as a tool to get 
insight into the non-perturbative dynamics. 




\section{AdS/CFT setup}

We will now describe how to set up an AdS/CFT computation for determining the 
spacetime behaviour of the energy-momentum tensor \cite{US1}. This 
method does not make any underlying assumptions about local equilibrium or 
hydrodynamical behavior. We will obtain hydrodynamic expansion as a generic late 
time behaviour of the expanding strongly coupled plasma. 

Suppose that we consider some macroscopic state of the plasma characterized by a 
spacetime profile of the energy-momentum tensor
\eq
T_{\mu\nu}(x^\rho)\ .
\eqx
Then, since the AdS/CFT correspondence asserts the exact equivalence of gauge 
and string theory, such a state should have its counterpart on the string side 
of the correspondence. Typically it will be given by a modification of the 
geometry of the original $AdS_5\times S^5$ metric. This comes from the fact that 
operators in gauge theory correspond to  fields in supergravity (or string 
theory). When we consider a state with a nonzero expectation value of an 
operator, the \emph{dual} gravity background will have the corresponding field 
modified from its `vacuum' $AdS_5 \times S^5$ value. In the case of the energy 
momentum tensor the corresponding field is just the 5-dimensional metric. One then has to 
assume that the geometry is well defined i.e. it does not have a naked 
singularity - a singularity not hidden by an event horizon -.
This principle will select the allowed physical spacetime profiles of 
gauge theory energy-momentum tensor. Thus together with the Einstein equations 
this becomes the main dynamical mechanism for the strongly coupled gauge theory.

The simplest way to formulate the precise correspondence between the expectation 
value of the energy-momentum tensor and bulk geometry is to use the 
Fefferman-Graham system of coordinates \cite{fg} for the latter:
\eq
ds^2=\f{g_{\mu\nu}(x^\rho,z) dx^\mu dx^\nu+dz^2}{z^2}\ .
\eqx 
This metric has to be a solution of 5-dimensional Einstein's equation with negative 
cosmological constant\footnote{One can show that such solutions lift to 10-dimensional 
solutions of ten dimensional type IIB supergravity. The effective 5-dimensional negative 
cosmological constant comes from the 5-form field in 10-dimensional supergravity.}:
\eq
\label{e.einst}
R_{\mu\nu}-\f{1}{2}g_{\mu\nu} R - 6\, g_{\mu\nu}=0\ .
\eqx
The expectation value of the energy momentum tensor may be easily recovered by 
expanding the metric near the boundary $z=0,$ following the ``holographic 
renormalization'' procedure \cite{Skenderis},
\eq
g_{\mu\nu}(x^\rho,z)=\eta_{\mu\nu}+z^4 g^{(4)}_{\mu\nu}(x^\rho)+\ldots\ .
\eqx
Then
\eq
\cor{T_{\mu\nu}(x^\rho)} = \f{N_c^2}{2\pi^2} \cdot g^{(4)}_{\mu\nu}(x^\rho)\ .
\eqx
This relation can be used in two ways. Firstly, given a solution of Einstein 
equations we may read off the corresponding gauge theoretical energy-momentum 
tensor. Secondly, given a traceless and conserved energy-momentum profile one 
may integrate Einstein equations into the bulk in order to obtain the dual 
geometry\footnote{{This can be done order by order in $z^{2}$, which is a 
near-boundary expansion. However potential singularities are hidden deep in the 
bulk, thus this power series needs to be resummed.}}. Then the criterion of 
nonsingularity of the geometry obtained in this way will determine the allowed 
spacetime evolution of the plasma. Let us note that this formulation is in fact 
quite far away from a conventional initial value problem.

Before we move to the case of expanding plasma, it is convenient to consider the 
simple situation of a static uniform plasma with a constant energy momentum 
tensor. Then the Einstein's equations can be solved analytically and we find 
\cite{US1}
that the exact dual geometry of such a system is
\eq
\label{e.bhfef}
ds^2=-\f{(1-z^4/z_0^4)^2}{(1+z^4/z_0^4)z^2}\ dt^2
+(1+z^4/z_0^4)\f{dx^2}{z^2}+ \f{dz^2}{z^2}\ .
\eqx 
This metric may look at first glance unfamiliar, but a change of coordinates
\eq
\zt=\f{z}{\sqrt{1+\f{z^4}{z_0^4}}}
\eqx
transforms it to the standard AdS Schwarzschild static black hole
\eq
ds^2=-\f{1-\zt^4/\zt_0^4}{\zt^2} dt^2
+\f{dx^2}{\zt^2}+\f{1}{1-\zt^4/\zt_0^4} \f{d\zt^2}{\zt^2}
\label{standard}\eqx
with $\zt_0=z_0/\sqrt{2}$ being the location of the horizon. Before we proceed 
further, let us note here one crucial thing: the fact, that the dual geometry of 
a gauge theory system with constant energy density is a black hole was {\em not} 
an assumption, but rather an outcome of a computation.

The Hawking temperature
\eq
T=\f{1}{\pi \zt_0} \equiv \f{\sqrt{2}}{\pi z_0}
\eqx
is then identified with the gauge theory temperature, and the entropy with the 
Bekenstein-Hawking black hole entropy 
\eq
S=\f{N_c^2}{2\pi\zt_0^3}=\f{\pi^2}{2} N_c^2 T^3
\eqx
which is $3/4$ of the entropy at zero coupling. To finish our discussion of the 
static black hole, we note that the Fefferman-Graham coordinates cover only the 
part of spacetime lying outside the horizon.

\section{Boost invariant flow}

Let us now apply the above procedure to a generic boost-invariant flow, in view of
making contact with the hydrodynamical Bjorken flow described 
in section 2. However we do not want to make any preassumptions on the dynamics, since
we would like to recover the hydrodynamic behaviour as an outcome of an AdS/CFT
computation. To this end let us consider the most general gauge theory energy-momentum tensor
which is boost-invariant and does not depend on transverse coordinates (see Fig.~1). Then
conservation of energy-momentum $\partial_\mu T^{\mu\nu}=0$ and tracelessness $T^\mu_\mu=0$ allow
to express all nonvanishing components of $T_{\mu\nu}$ in terms of a single 
function $\eps(\tau)$ -- the energy density at mid rapidity:
\eq
\label{e.tgen}
T_{\mu\nu}\! = \!
\left(\begin{tabular}{cccc}
$\eps(\tau)$ & 0 &0 & 0 \\
0 & $-\tau^3 \f{d}{d\tau} \eps(\tau)\!-\!\tau^2 \eps(\tau)$ & 0 & 0 \\
0 & 0 & $\eps(\tau)\!+\! \f{1}{2}\tau \f{d}{d\tau} \eps(\tau)$ & 0 \\
0 & 0 & 0 & $\eps(\tau)\!+\! \f{1}{2}\tau \f{d}{d\tau} \eps(\tau)$
\end{tabular}\right)
\eqx
\smallskip

Let us concentrate, 
following \cite{US1} on the late time asymptotics of this function i.e.
\eq
\eps(\tau) \sim \f{1}{\tau^s} +\ldots
\eqx
for $\tau \to \infty$. Energy positivity requires that $0 \leq s \leq 4$. We 
will consider sharp inequalities here\footnote{Recently the case $s=4$ has been 
considered in \cite{Kajantie:2008jz}.}. The most general metric consistent with 
the symmetry assumptions is
\eq
\label{e.ansatz}
ds^2=\f{-e^{a(\tau,z)} d\tau^2 +\tau^2 e^{b(\tau,z)} dy^2
  +e^{c(\tau,z)} dx^2_\perp}{z^2} +\f{dz^2}{z^2}  \ .
\eqx
In order to find the late time form of the solution corresponding to 
$\eps(\tau)=1/\tau^s$ we may solve the Einstein equations in a power series for 
the metric coefficients
\eq
\label{e.aexp}
a(\tau,z)=\sum_{n=0}^N a_n(\tau) z^{4+2n}
\eqx
where $a_0(\tau)=-\eps(\tau)=-1/\tau^s$. Then from each coefficient $a_n(\tau)$ 
we may extract the leading large $\tau$ behaviour and neglect the subleading 
terms. It turns out that this procedure is exactly equivalent to introducing a 
scaling variable
\eq
v=\f{z}{\tau^{\f{s}{4}}}
\eqx
and assuming the metric coefficients to be just functions of $v$ e.g. 
$a(z,\tau)=a(v)$ in the large proper time limit (namely $\tau \to \infty$, $z\to 
\infty$ with $v$ kept fixed). In this limit Einstein's equations become just 
ordinary differential equations and may be solved analytically. 
The singularity of these geometries can then be tested by computing the scalar 
curvature invariant
\eq
\rsq=R^{\mu\nu\alpha\beta}R_{\mu\nu\alpha\beta}\ .
\eqx
Since our solutions are defined only in the large proper time limit $\tau \to 
\infty$ with the scaling variable $v$ kept fixed, we have to evaluate $\rsq$ in 
the same manner\footnote{{It should be stressed however, that this condition is 
really a condition of regularity of the expansion of the curvature invariant. It 
is safe to do as long as each term in the large proper-time expansion is 
regular. On the other hand any singularity present in this expansion might be 
either a genuine curvature singularity or a singularity of the expansion, 
see a detailed discussion in  section \ref{secsing}.}}.

This procedure is described in detail in \cite{US1}. The result is that
\begin{itemize}
\item for generic $s$ the resulting solution is singular
\item the only nonsingular solution corresponds to $s=\f{4}{3}$ which is just 
the hydrodynamic Bjorken expansion (see \eqref{g3}, section 2)
\item the resulting metric takes the form
\end{itemize}
\eqn
\label{e.flgeom}
ds^2=\f{1}{z^2} \left[- \f{\left( 1-\f{e_0}{3}
      \f{z^4}{\tau^{4/3}}\right)^2}{1+\f{e_0}{3}\f{z^4}{\tau^{4/3}}} d\tau^2+
\left( {\textstyle 1+\f{e_0}{3} \f{z^4}{\tau^{4/3}}}\right) (\tau^2
      dy^2 +dx^2_\perp)\right] 
+ \f{dz^2} {z^2}
\eqnx
where we reinstated the dimensionful parameter $e_0$ so that
\eq
\eps(\tau)=e_0/\tau^{\f{4}{3}}\ .
\eqx 
Let us note some salient features of this result. The geometry (\ref{e.flgeom}) 
bears striking resemblance to the AdS black hole geometry (\ref{e.bhfef}) but 
with the position of the `effective horizon' being time dependent
\eq
z_0=\sqrt[4]{\f{3}{e_0}} \cdot \tau^{\f{1}{3}}
\eqx
Then assuming similar relations as for the black hole case one gets the Bjorken 
scaling of the temperature and entropy.
\eqn
T &=& \f{\sqrt{2}}{\pi z_0}= \f{2^{\f{1}{2}} e_0^{\f{1}{4}}}{\pi 3^{\f{1}{4}}} 
\tau^{-\f{1}{3}}\nonumber\\
S &\propto& \frac{\tau}{z_0^3} =const\ .
\eqnx
We see that the `movement' of the horizon into the bulk of AdS corresponds 
physically to cooling of the expanding gauge theory plasma system.

A significant fact that has to be kept in mind is that the geometry 
(\ref{e.flgeom}), in contrast to (\ref{e.bhfef}), is not an exact solution of 
Einstein's equation. It is valid only for large times. For smaller times it has 
to be modified. We will now discuss this issue in more detail as it reflects 
important physical properties of the gauge theory plasma.

\section{Beyond perfect fluid}

The geometry (\ref{e.flgeom}) is only a solution of Einstein equations in the 
scaling limit. However with some effort, one can get also the first subleading 
corrections to the metric i.e.
\eq
\label{e.flexp}
a(z,\tau)=a_0(v)+\f{1}{\tau^{\f{4}{3}}}\ a_2(v) +\ldots
\eqx
Then after evaluating $\rsq$, keeping track of subleading terms we find
\eq
\rsq=R_0(v)+ \f{1}{\tau^{\f{4}{3}}} R_2(v) +\ldots
\eqx
where $R_0(v)$ is finite, but $R_2(v)$ develops a $4^{th}$ order pole 
singularity. The physical meaning of this behaviour is indeed quite clear. The 
geometry (\ref{e.flexp}) is dual to a state in gauge theory which undergoes 
expansion according to exact {\em perfect fluid} hydrodynamics. Yet we know that 
gauge theory plasma has nonzero viscosity and hence the perfect fluid behaviour 
\eq
\eps(\tau)=\f{1}{\tau^{\f{4}{3}}}
\eqx
is not exact but, if it would be described by viscous Bjorken expansion (viscous 
hydrodynamics), it would be modified to
\eq
\eps(\tau)=\f{1}{\tau^{\f{4}{3}}}\left( 1-\f{2\eta_0}{\tau^{ \f{2}{3}}} +\ldots 
\right)
\eqx
where $\eta_0$ is related to the shear viscosity through $\eta=\eta_0/\tau$ 
(which follows from the scaling $\eta \propto T^3$).

Let us show how this arises using the AdS/CFT methods.
We will not presuppose a specific form of subleading correction but will start 
from
\eq
\eps(\tau)=\f{1}{\tau^{\f{4}{3}}}\left( 1-\f{2\eta_0}{\tau^r} +\ldots \right)
\eqx 
with a generic $r$. In order to verify that plasma expansion follows viscous 
hydrodynamics we will have to first show that $r=\f{2}{3}$. The metric 
coefficients will now have an additional piece scaling as $\f{1}{\tau^{r}} 
a_r(v)$. It turns out that the curvature scalar $\rsq$ is always {\em 
nonsingular} at that order\footnote{This was first observed for $r=2/3$ in 
\cite{Nak1}.}. Hence we have to go one order further i.e. find all coefficients 
appearing in the following expansion
\eq
a(z,\tau)=a_0(v)+\f{1}{\tau^{r}} a_r(v)+\f{1}{\tau^{2r}} a_{2r}(v)+ 
\f{1}{\tau^{\f{4}{3}}} a_2(v) +\ldots
\eqx
Then the curvature scalar has the form
\eq
\rsq=R_0(v)+\f{1}{\tau^{r}} R_r(v)+\f{1}{\tau^{2r}} R_{2r}(v)+ 
\f{1}{\tau^{\f{4}{3}}} R_2(v) +\ldots
\eqx
with $R_0(v)$ and $R_r(v)$ being nonsingular, while {\em both} $R_{2r}(v)$ and 
$R_2(v)$ turn out to have $4^{th}$ order pole singularities. In order for them 
to have a chance to cancel we have to have
\eq
r=\f{2}{3}
\eqx
which is exactly the scaling of a viscosity correction to Bjorken flow. Moreover 
cancelation occurs only when the shear viscosity coefficient has the 
value\footnote{We set here $e_0=1$.}
\eq
\eta_0=2^{-\f{1}{2}} 3^{-\f{3}{4}}
\eqx
which is equivalent to $\eta/s=1/4\pi$ (for details see \cite{RJ}). In a similar 
manner one can go one order higher and determine a coefficient of second order 
hydrodynamics. However at that order, it turns out that there remains a leftover 
logarithmic singularity. We will show, in section \ref{secsing}, that the 
logarithmic singularity arises due to a pathology of the Fefferman-Graham 
expansion and can be  avoided when one makes a different late time expansion.

Finally let us comment on why it is interesting to verify the viscous 
hydrodynamic behaviour with the specific viscosity coefficient for the expanding 
plasma. Already before, there have been studies of {\em linearized} perturbations 
around the uniform plasma which demonstrated that hydrodynamic behaviour appears 
for small fluctuations and the value of viscosity was obtained from the Kubo 
formula 
\cite{son}. It was interesting to verify whether hydrodynamics also applies in 
its fully 
nonlinear regime. The agreement of the resulting value of the viscosity 
coefficient is thus a nontrivial consistency check.

Another motivation for developing an AdS/CFT framework for studying such 
time-dependent phenomenae is the fact that some of the most interesting and 
puzzling phenomena in heavy ion collisions are definitely very far from 
equilibruum. We will mention some examples in section \ref{seclast}.

\section{Beyond boost-invariance}

The calculations presented in the previous sections were performed for systems 
with boost invariance symmetry and full translational and rotational symmetry in 
the transverse plane. This allowed us to perform explicit computations as the 
symmetry assumptions effectively reduced the calculation to systems of ordinary 
differential equations. 
In this manner we obtained directly the solution for gauge theory energy density 
$\eps(\tau)$. Then, in order to find the link with hydrodynamics, we found that 
this solution is a solution of hydrodynamic equations with specific values for 
the transport coefficients. 

This approach has both an advantage and a drawback. The advantage is that one 
does not presuppose any kind of dynamics and one may try to apply it in contexts 
very far from equillibrum, where hydrodynamic description does not apply. The 
drawback is that the appearance of hydrodynamic equations is not transparent and 
it is difficult to relax the symmetry assumptions due to the complexity of solving 
nonlinear Einstein's equations.

Recently the latter drawback was addressed and it was shown in general how the 
equations of hydrodynamics arise from the gravity side \cite{MINWALLA}. Here we 
will briefly review this approach.

Let us start from the static black hole (\ref{standard},\ref{e.bhfef}) but 
written in yet 
another coordinate system -- the incoming Eddington-Finkelstein coordinates:
\eq
ds^2=-2dt dr-r^2 \left(1- \f{T^4}{\pi^4 r^4} \right)dt^2 +r^2\eta_{ij}  dx^i 
dx^j\ .
\eqx
Here $T$ is the temperature, $r=\infty$ corresponds to the boundary. 
$x^\mu=const$ are null curves going from the boundary into the black hole. The 
advantage of this coordinate system is that it is well defined on the horizon 
and extends all the way from the boundary to the singularity at the center of 
the black hole.

The geometry given above corresponds to a uniform plasma at rest (i.e. with the 
4-velocity $u^\mu=(1,0,0,0)$) and given temperature $T$. We may now perform a 
boost (and perform a dilatation) to obtain the dual geometry to a moving plasma system 
with uniform 4-velocity $u^\mu$ and temperature $T$:
\eq
\label{e.boosted}
ds^2=-2u_\mu dx^\mu dr-r^2 \left(1- \f{T^4}{\pi^4 r^4} \right)u_\mu
u_\nu dx^\mu dx^\nu +r^2(\eta_{\mu \nu}+u_\mu u_\nu)  dx^\mu dx^\nu
\eqx
The idea of ref.\cite{MINWALLA} is to allow $u^\mu$ and $T$ to be 
(slowly-varying) functions of the spacetime coordinates. Once this is done the 
geometry  (\ref{e.boosted}) ceases to be an exact solution of Einstein equation 
because of nonvanishing gradients of the parameters $u^\mu$ and $T$. This 
suggests to perform an expansion of the solution in terms of gradients which has 
been carried out in \cite{MINWALLA} up to second order in derivatives. The 
integration constants arising at each order are again fixed by requiring 
regularity of the metric at the horizon. The resulting metric is expressed in 
terms of 4-velocities and temperatures and their derivatives, so when one 
extracts the energy-momentum tensor it will be given directly in terms of those 
quantities. Up to first order the expression is
\eq
\label{e.tmunumin}
T^{\mu\nu}=\f{N_c^2}{8\pi^2} \left\{(\pi T)^4 (\eta^{\mu\nu}+4u^\mu u^\nu)- 
2(\pi T)^3 \sg_{shear}^{\mu\nu} \right\} \ .
\eqx
The first term is just the perfect fluid energy momentum tensor, while the 
second term involves the shear viscosity. 
This result essentially demonstrates how general hydrodynamic equations arise 
from gravity in AdS/CFT. Indeed, once it is shown that the general form of the gauge 
theory energy-momentum tensor has the form (\ref{e.tmunumin}), then conservation 
of energy momentum $\partial_\mu T^{\mu\nu}=0$ is equivalent, by definition, to conformal 
relativistic Navier-Stokes equations. 
As a byproduct, the above construction also gives a map from solutions of 
(viscous) hydrodynamics to gravity solutions. However this setup requires that 
the starting point is not far off from equilibruum. For processes which do not 
admit a hydrodynamic description (like the early stage of a heavy-ion collision) 
one has to resort to different methods.

\section{Reduction of Singularities}
\label{secsing}

The leftover logarithmic singularity found in the third order of the square of 
the Riemann tensor \cite{Heller:2007qt} (as well as in the higher curvature 
invariants \cite{Benincasa:2007tp}) might be the signal of either genuine 
curvature singularity or the singularity of the chosen expansion 
scheme\footnote{As it was stressed before, the large-proper time expansion of 
curvature invariants is not diffeomorphism-invariant. The encountered 
singularities are 
physical only if there is no coordinate transformation which removes them.}. If 
the first is true, this means that the whole framework is inconsistent and 
either one needs to include additional degrees of freedom to cure it or the 
boost-invariant flow is unphysical\footnote{Since it corresponds to the naked 
singularity on the gravity side}. The results presented in 
\cite{Benincasa:2007tp} show that no supergravity field can fix the problem, 
which led to conjectures, that boost-invariant flow cannot be realized within the
supergravity framework \cite{Buchel:2008xr}. On the other hand, the gravity dual of 
general fluid flow up to the second order in derivatives was shown to be regular 
and it was hard to imagine how possible singularities could form in the third 
order \cite{MINWALLA, Bhattacharyya:2008xc}. The resolution of this puzzle was 
presented in \cite{Heller:2008mb} (see also \cite{Kinoshita:2008dq}). It turns 
out that there exists a singular coordinate transformation from Fefferman-Graham 
coordinates to Eddington-Finkelstein ones, which yields a completely regular and 
smooth metric from the boundary up to the standard black-brane singularity. This 
leads to regular (apart from the standard black-brane singularity) large proper-time 
expansion of curvature invariants. The metric ansatz in Eddington-Finelstein 
coordinates takes the form

\eq
\mathrm{d}s^{2} = 2 \mathrm{d} \tilde{\tau} \cdot \mathrm{d} r - r^{2} \tilde{A} 
\left( \tilde{\tau}, r \right) \mathrm{d} \tilde{\tau}^{2} + \left( 1 + r \cdot 
\tilde{\tau} \right)^{2} e^{\tilde{b} \left( \tilde{\tau}, r \right)} \mathrm{d} 
y^{2} + r^{2} e^{\tilde{c} \left( \tilde{\tau}, r\right)} \mathrm{d} x_{\perp}^{2} 
\eqx

\noindent and was motivated by the boosted black-brane metric (\ref{e.boosted}) 
with a boost and dilatation parameters $u = 1 \cdot \partial_{\tilde{\tau}}$ and 
$T \sim \tilde{\tau}^{-4/3}$. The functions $\tilde{A} \left( \tilde{\tau}, r 
\right)$, $\tilde{b} \left( \tilde{\tau}, r \right)$ and $\tilde{c} \left( 
\tilde{\tau}, r\right)$ are expanded in a large-proper time expansion 
analogously as it was in the Fefferman Graham case, i. e.

\eq
\tilde{A} \left( \tilde{\tau}, r \right) = \tilde{A}_{0} \left( r \cdot 
\tilde{\tau}^{1/3} \right) + \frac{1}{\tilde{\tau}^{2/3}} \tilde{A}_{1} \left(  
r \cdot \tilde{\tau}^{1/3} \right) + \frac{1}{\tilde{\tau}^{4/3}} \tilde{A}_{2} 
\left(  r \cdot \tilde{\tau}^{1/3} \right) + \ldots
\eqx

\noindent This form of expansion can also be justified by 
\cite{Bhattacharyya:2008ji}. The terms damped by inverse power of proper time come 
from the gradient expansion. The boundary metric in proper-time-rapidity 
coordinates has non-vanishing Christoffel symbols $\Gamma \sim \ttau^{-1}$, thus 
the four velocity gradient $\nabla u$ (which is constant in these 
coordinates) gives a factor of $\ttau^{-1}$. On the other hand the expansion 
parameter multiplying each term in gradient expansion is the inverse power of 
the temperature $T$(see \cite{MINWALLA}). Because $T \sim 
\ttau^{-1/3}$, the overall damping is indeed $\ttau^{-2/3}$ - a fact derived in 
\cite{Heller:2007qt} from the non-singularity argument.

 The 
non-perturbative\footnote{In the sense of large-proper time expansion} piece in 
the metric at $\mathrm{d} y^{2}$ introduced in \cite{Kinoshita:2008dq} is 
responsible for a correct limit energy density $\rightarrow 0$. It also becomes 
important if one wants to solve the problem of early-time dynamics 
\cite{Kovchegov:2007pq} using Eddington-Finkelstein coordinates.

The integration 
constants\footnote{Not all of them -- there is a remaining gauge freedom 
(coordinate transformation), which leaves the metric ansatz unchanged: $r 
\rightarrow r + f\left( \tilde{\tau} \right)$, where $f\left( \tilde{\tau} 
\right)$ is an arbitrary function} are fixed by requiring the regularity of the 
metric functions $\tilde{A}_{i} \left( \tilde{v} \right)$, $\tilde{b}_{i} \left( 
\tilde{v} \right)$ and $\tilde{c}_{i} \left( \tilde{v} \right)$ at each order 
$i$. This is justified since the Eddington-Finkelstein are valid for 
$\tilde{\tau} > 0$ and $0 < \tilde{v} = r \cdot \tilde{\tau} < \infty$. The 
singular coordinate transformation which takes the metric from 
Eddington-Finkelstein coordinates to Fefferman-Graham ones is given order by 
order in the gradient expansion by

\eqn
\tilde{\tau} \left( \tau, z \right) &=& \tau \cdot \Big\{ T_{0} \left(z \cdot 
\tau^{-1/3} \right) + \frac{1}{\tau^{2/3}} T_{1} \left(z \cdot \tau^{-1/3} 
\right) + \ldots \Big\} \ \mathrm{,}\\
r \left( \tau, z \right) &=& \frac{1}{z} \cdot \Big\{ R_{0} \left(z \cdot 
\tau^{-1/3} \right) + \frac{1}{\tau^{2/3}} R_{1} \left(z \cdot \tau^{-1/3} 
\right) + \ldots \Big\}\ \mathrm{.}
\eqnx
The leading-order solutions (corresponding to the perfect fluid on the gauge 
theory side) 
 are related by
\eqn
\label{coordtrafo}
\ttau &\rightarrow& \tau \left\{ 1 - \frac{1}{\tau^{2/3}} \left[ \frac{3^{1/4}
    \pi}{4 \sqrt{2}} + \frac{3^{1/4}}{2 \sqrt{2}}
  \tan^{-1}{\left(\frac{3^{1/4}}{\sqrt{2}} r \cdot \tau^{1/3}\right)}\  + 
\right.\right. \nonumber\\ &&
\ \ \ \ \ \ \ \ \ \ \ \ \ \ \ \ \ \ \ \ \ + \left.\left.\frac{3^{1/4}}{4 
\sqrt{2}} \log {\left(\frac{r \cdot \tau^{1/3} -
      \frac{\sqrt{2}}{3^{1/4}}}{ r \cdot \tau^{1/3} +
      \frac{\sqrt{2}}{3^{1/4}}}\right)} \right] \right\}\ \mathrm{,}
\nonumber \\ 
r &\rightarrow& \frac{1}{z} \cdot \sqrt{1 + \frac{z^{4}}{3 \cdot \tau^{4/3}}}
\mathrm{.} 
\eqnx
The transformation is singular at $z = 3^{1/4} \tau^{1/3}$, which is precisely 
the locus of the logarithmic singularity encountered in \cite{Heller:2007qt}. 
Formulas for higher order transformation coefficients are too long to be 
presented here and can be found in \cite{Math}. 
The energy-momentum tensor extracted from the solution in Eddington-Finkelstein 
coordinates reproduces the energy momentum tensor obtained in 
\cite{Heller:2007qt}.

\section{Beyond hydrodynamics}

\label{seclast}

Gauge-gravity duality has already proven to be an invaluable tool in describing 
properties of static or near-equilibrium (hydrodynamics) strongly coupled gauge 
theory systems. Noticeable achievements in that direction are the viscosity 
evaluation bound 
\cite{son} and the consistent formulation of the second order conformal 
hydrodynamics \cite{Baier:2007ix, MINWALLA}. These successes came 
from the holographic understanding of hydrodynamics. On the other hand there is 
much more interesting and nontrivial dynamics than hydro. Far from equilibrium 
behavior of gauge theories is a fascinating and pretty much open problem of 
experimental importance, like the early universe or initial stages of heavy ion 
collisions\footnote{In the late stages of heavy ion collision, strongly coupled 
quark-gluon plasma forms and holographic technics at strong gauge coupling are 
better justified then just after the collision (running of the coupling). 
Nevertheless applying AdS/CFT correspondence to describe far from equilibrium 
processes in gauge theories is an interesting problem even from a purely 
theoretical point of view.}. The AdS/CFT correspondence is surely capable to  
shed new light on these problems, or even be understood as a formulation of far 
from equilibrium gauge theory.

In the context of heavy-ion collisions the most important and probably the most 
difficult questions concern the issues of early time dynamics 
\cite{Kovchegov:2007pq} and the transition to an isotropic \cite{Janik:2008tc} 
and 
thermalized medium. One of the puzzles here is the short time in which nuclear 
matter approach local equilibrium. Experimental data fitting well with 
hydrodynamical simulations with small viscosity justified applications of the 
AdS/CFT correspondence at strong coupling for the late stages of heavy-ion 
collisions. It is not clear to what extent early time dynamics is driven by 
non-perturbative effects and whether the  lessons learned from AdS/CFT might be 
directly applied to the nuclear matter in the early stages of the evolution. 
Approaching equilibrium is also of an interest from the General Relativity point 
of view. Isotropic 
and thermalized matter on the gauge theory side corresponds to a black hole in 
AdS, whereas thermalization and approach to local equilibrium should be governed 
by the dynamics of gravitational collapse.

Perhaps some of these questions might be answered by studying collisions of 
shock-waves in AdS. The geometry corresponding to a projectile in 3+1 dimensions 
was constructed in \cite{US1} using holographic renormalization. The 
metric
\eq
\mathrm{d} s^{2} = \frac1{z^{2}}\left\{- 2 \mathrm{d} x^{+} \mathrm{d} x^{-} + 
f\left( x^{-} 
\right) \cdot z^{4} \left( \mathrm{d} x^{-} \right)^{2} + \mathrm{d} 
x_{\perp}^{2} + \mathrm{d} z^{2}\right\}
\eqx
with an arbitrary function $f \left( x^{-} \right)$ corresponds to the situation 
when

\begin{itemize}
\item the dynamics is one-dimensional (i.e. no dependence on transverse coordinates),
\item the energy-momentum tensor depends only on a single light-cone variable (here 
chosen to be $x^{-} = x^{0} - x^{1}$).
\end{itemize}

Traceless and conserved energy momentum tensor satisfying the above assumptions 
takes a particularly simple form -- its only non-zero component is $T^{- -}  = f 
\left( x^{-} \right)$. Choosing $f \left( x^{-} \right) = M \delta \left( x^{-} 
\right) $ leads to a shock-wave -- infinitely thin plane of matter moving at the 
speed of light, which is a toy-model for highly boosted nucleus. The idea is to 
collide two such projectiles and single out the physical behavior of the plasma 
from the regularity of the dual geometry. This is a difficult problem, because 
of the broken boost-invariance, which leads to solving Einstein equations in 3 
variables ($x^{+}$, $x^{-}$ and $z$ or equivalently $\tau$, $y$ and $z$). The 
geometry before the collision ($x^{+} + x^{-}  <  0$) is known -- it is simply 
the superposition of two incoming shock-waves

\eq
\mathrm{d} s^{2} = \frac1{z^{2}}\left\{- 2 \mathrm{d} x^{+} \mathrm{d} x^{-}\! + 
\! M  \delta 
\left( x^{-} \right) z^{4} \left( \mathrm{d} x^{-} \right)^{2} \!+\! M 
 \delta \left( x^{+} \right) z^{4}  \left( \mathrm{d} x^{+} 
\right)^{2} \! +\! \mathrm{d} x_{\perp}^{2} \!+ \!\mathrm{d} z^{2}\right\} 
\mathrm{.}
\eqx

Shock-waves collide at $x^{+} = x^{-} = 0$ and from now on the dynamics of the 
system must be deduced from Einstein equations. The first attempt to address 
this issue in \cite{Kajantie:2008rx}
focused on the simpler setup than presented so far -- a shock-waves collision in 
1+1 dimensions. The energy-momentum tensor for such a system before the 
collision is given by $T_{+ +} = f \left( x^{-} \right) $ and $T_{- -} = g 
\left( x^{+} \right) $ with vanishing off-diagonal components. This is at the 
same time the most general form of the energy-momentum tensor for a 1 + 1 
dimensional CFT. A nice feature is that the dual geometry for the {\em whole} 
collision process can be constructed here exactly. However in this low 
dimensional context, the projectiles pass each other unaffected (or propelled  
back-to-back \cite{Kajantie:2008rx}), so the physics 
of plasma production and thermalization is  absent here. On the 
other hand the problem of genuine interest -- collision of shock-waves in 3+1 
dimensions -- requires some approximation scheme in which Einstein equations 
become tractable. Up to now, two proposals have been made. The first one 
\cite{Grumiller:2008va} treats proper time as a small parameter but suffers from 
a negative energy density in some regions due to the conditions imposed at the 
light-cone. The second one \cite{Albacete:2008vs} solves Einstein equations 
perturbatively in $M$ leading to the prediction that shock-waves stop almost 
immediately after the collision (reminding of the {\it full stopping} condition 
of the Landau flow, see section \ref{sechydro}.). This problems surely deserve further 
studies. 

There are also other studies of dynamical processes in an evolving plasma system 
which go beyond hydrodynamics. One example is the problem of thermalization of 
small perturbations around the expanding plasma (some first investigations has 
been performed in \cite{US1}). Another use of the evolving geometries 
is to study other physical processes in the presence of the evolving plasma 
system like e.g. the physics of mesons and flavours studied through embedding D7 
branes in the time-dependent geometries \cite{flavors}. Finally one may study 
isotropisation of anisotropic plasma. The first investigations have been 
performed in \cite{Janik:2008tc} (see the talk by P. Witaszczyk in the same 
proceedings).

\section{Summary}
The Gauge/Gravity approach to the formation and evolution of a quark-gluon 
plasma in heavy-ion collisions described above has the interest of casting an 
exploratory bridge between the rigorous results of string theory and some 
pending  questions raised by the experiments on quark-gluon plasma. These 
questions cannot yet be raised in the framework of strongly coupled QCD, for 
which we do not possess the adequate tools, but they can be addressed for the 
first time in a quantitative and rigorous way in the supersymmetric case of the 
AdS/CFT correspondence. It is thus a novel and valuable approach and can serve 
as a model for further studies.

Let us summarize some aspects of this approach, being aware (and with apologies 
for those not quoted or mentioned) that this subject is in constant development which
will force us to mention only a few of them.

Starting with the experimental evidence that hydrodynamics is relevant in the 
formation and evolution of a quark-gluon plasma in heavy-ion collisions and in 
particular  of the ``Bjorken flow'' description, we present the AdS/CFT setup 
allowing to describe the dynamics of the plasma (in the AdS/CFT case). We show 
that it is possible to derive the geometry dual to the asymptotic evolution of 
the plasma in terms of an expansion in a scaling variable. the nonsingularity 
requirement on the Gravity side gives a set of ``selection rules'' on the Gauge 
side: the perfect fluid at first order, the $\eta/s=1/4\pi$ property and other 
transport coefficients at higher orders...

Beyond the boost-invariant Bjorken flow, there exists an intriguing but rigorous 
one-to-one correspondence between the complete (and even completed using 
AdS/CFT!) hydrodynamic equations and the solutions of the Einstein equations in 
the bulk of the 5-dimensional space. Some apparent obstacles, such as the appearance of 
logarithmic singularities in the asymptotic expansion of the 5-dimensional metric, have been 
shown to be a mere artefact of the choice of the expansion parameter and have 
been cured.

Beyond the hydrodynamical description of the transient plasma, one goal is now 
to exoplore the dynamical aspects far from equilibrium. We describe some very 
recent attempts, which even though not conclusive yet, show the interest of the 
extension of  Gauge/Gravity correspondence to attack some down-to-earth 
problems, such as the short thermalization time, probably observed by the 
heavy-ion phenomenology and more generally the effect of the initial conditions 
on the whole process.

It is quite interesting  to see that some complex aspects of Gauge field theory 
dynamics can find unexpected answers from Gravity. It would  be intringuing that 
some nontrivial aspects of Gravity (such as the dynamics of moving black holes) 
also could gain some new insight from the correspondence with some aspects of 
heavy-ion collisions.

\section*{Acknowledgements}
This investigation was partly supported by the     
MEiN research grants 1 P03B 04529 (2005-2008) (RP), 1 P03B 04029 (2005-2008) (RAJ and MPH) and N202 247135 (2008-2010) (MPH), by 6 Program of European
Union ``Marie Curie Transfer of Knowledge'' Project: Correlations in Complex
Systems ``COCOS'' MTKD-CT-2004-517186 and the RTN network ENRAGE MRTN-CT-2004-005616. MPH also acknowledges the support from the British Council within the Polish-British Young Scientists Programme and hospitality of Durham University while these notes were being completed.

\end{document}